\documentclass[12pt]{iopart}
\usepackage{epsfig,float,iopams,setstack}
\usepackage{graphicx}

\usepackage{latexsym}
\usepackage{epstopdf}
\usepackage{amssymb}

\def\of{\omega_{\phi}}

\DeclareMathAlphabet{\mathsfsl}{OT1}{cmss}{m}{sl}

\def\R{\hbox{{\rm I}\kern-0.2em{\rm R}\kern0.2em}}

\def\bn{\begin{equation}}
\def\en{\end{equation}}
\def\bny{\begin{eqnarray}}
\def\eny{\end{eqnarray}}
\def\be{\begin{eqnarray*}}
\def\ee{\end{eqnarray*}}
\def\bc{\begin{center}}
\def\ec{\end{center}}

\def\({\left(}
\def\){\right  )}
\def\[{\left[}
\def\]{\right]}
\def\bc{\begin{center}}
\def\ec{\end{center}}

\begin{document}

\title[]{\bf \Large Accretion on Reissner-Nordstr\"{o}m-(anti)-de Sitter Black Hole with Global Monopole}

\author{Ayyesha K. Ahmed$^1$, Ugur Camci$^2$, and Mubasher Jamil$^1$}

\address{$^1$Department of Mathematics, School of Natural Sciences (SNS), National University
of Sciences and Technology (NUST), H-12, Islamabad, Pakistan}

\address{$^2$Department of Physics, Faculty of Sciences,
Akdeniz University, 07058, Antalya, Turkey}

\ead{ayyesha.kanwal@sns.nust.edu.pk, ucamci@akdeniz.edu.tr and mjamil@sns.nust.edu.pk}

\begin{abstract}
In this paper, we investigate the accretion on the Reissner-Nordstr\"{o}m anti-de-Sitter black hole with global monopole charge. We discuss the general solutions of accretion using the isothermal and polytropic equations of state for steady state, spherically symmetric, non-rotating accretion on the black hole. In the case of isothermal flow, we consider some specific fluids and derive their solutions at the sonic point as well. However, in case of polytropic fluid we calculate the general expressions only, as there exists no global (Bondi) solutions for polytropic test fluids. In addition to this, the effect of fluid on the mass accretion rate are also studied. Moreover, the large monopole parameter $\beta$ greatly suppresses the maximum accretion rate. 
\end{abstract}


\qquad \qquad \today


\section{Introduction}\label{INT}
In astrophysics, the accretion of fluids or matter onto compact objects such as neutron stars or black holes, is an interesting physical process as it describes a scenario which is most likely to explain the high energy output from the active galactic nuclei and quasars. Accretion is the process of capturing the matter by a gravitating object towards its center which leads to increase the mass and angular momentum of the accreting body. The stars and planets are formed by the process of accretion in dust clouds. The existence of supermassive black holes at the center of galaxies suggests that such black holes could have been gradually developed through the accretion process. An accretion disk is developed when dust and gases rotate around a compact object and accumulate into a disk. However, accretion does not always increase the mass of the accreting body but it could also decrease the mass such as accretion of phantom energy \cite{M1,M2}. A unique feature of black hole is the presence of an event horizon which acts as a boundary through which infalling fluid disappears. This may have various implications. For instance, it provides the inner boundary condition which describes the motion of the fluid and further it helps to avoid the uncertainties regarding the correct boundary conditions.

\par In $1952$ Bondi investigated the accretion process for a star within the Newtonian framework \cite{Bondi}. After the evolution of the relativistic theory of gravitation given by Einstein, it became possible for the astrophysicists to study the accretion process within the relativistic framework. So, Michel was the first one, who investigated the accretion process onto the Schwarzschild black hole \cite{Michel}. Further, Shapiro and Teukolsky \cite{Shapiro} also contributed to Michel's idea. Later on, Babichev et al.\cite{Babichev} discussed the effect of accretion of a dark energy onto Schwarzschild black hole and found that accretion of phantom energy will decrease the mass of the black hole. Debnath \cite{Debnath} further generalized the Babichev's idea and presented a framework of a static accretion onto general static and spherically symmetric black holes. Moreover in the series of recent papers \cite{1t,t1,PMach,Mjamil} accretion of a spherically symmetric spacetime with is investigated. The authors were mainly interested to determine that how the presence of the cosmological constant $\Lambda$ effects the accretion rate.

\par During the phase transition, many topological defects are produced such as cosmic strings, domain walls, monopoles, etc. Monopoles are the three-dimensional topological defects that are formed when the spherical symmetry is broken during the phase transition. These monopoles have Goldstone fields with high energy density that decrease with the distance only as $r^{-2}$, so that the total energy density is divergent at the large distances \cite{Bar}. This large density suggests that global monopoles can produce the strong gravitational fields. Another interesting point is that the spacetime around a cosmic string is conical globally but locally the spacetime is flat i.e. just a Minkowski spacetime. So we can think about this spacetime as Minkowski spacetime but, in case of a global monopole the spacetime is not locally flat. However, globally the spacetime at the equatorial plane is conical having deficit angle as $\Delta=8\pi^{2}\eta^{2}$ \cite{Jusufi}. Hence, it is evident that these monopoles exerts practically no gravitational force on non-relativistic matter but the space around it has a deficit solid angle and all the light rays are deflected by that angle, independent of the impact parameter \cite{Bar}. It is believed that the charged black holes do not exist in nature, since initial charge will be neutralized quickly \cite{Punsly}. However, these charged black holes may be produced during the gravitational collapse of massive stars particularly magnetized rotating stars surrounded by a magnetosphere having an equal and opposite charge. The magnetosphere preserves the black hole from a neutralization due to the accretion of charge, so that the black hole remains stable in a typical astrophysical environment having low density.

\par Barriola and Vilenkin found the approximate solution of the Einstein equations for the static spherically symmetric black hole with a global monopole \cite{Bar}. Several other authors have also studied the physical properties of the black holes with global monopoles \cite{Bronnikov,Hyu,Pitelli,Rahaman}.  In the absence of the electric charge $Q$ and the cosmological constant $\Lambda$, Letelier \cite{Letelier} obtained the metric representing the black hole spacetime with a spherical mass $M$ centered at the origin of the system of coordinates, surrounded by a spherical cloud of strings. The accretion process onto the black hole with a string cloud parameter, which can be interpreted as a global monopole, is examined by Ganguly et al.\cite{ggm}. The circular geodesics and accretion disk of a black hole with global monopole is discussed by Sun et al \cite{Sun}. Recently, the accretion process of a charged black hole in the anti-de-Sitter spacetime is investigated by Ficek \cite{Ficek}. The author calculated the analytical solutions for the isothermal and polytropic fluids and found a way to find sonic points as critical points of a Hamiltonian system. He has also proved the existence of closed trajectories and therefore homoclinic solutions in the phase space of subrelativistic isothermal flows. Here, we extend his idea by considering a global monopole charge $\beta$, which is different from the electric charge $Q$ in the black hole. In this paper we investigate the spherically symmetric, steady flow of the most general perfect fluid (gas). The basic motivation comes from the fact that some features of Reissner-Nordstr\"{o}m black hole resemble with the Kerr solution as they both share the similar structure of horizons \cite{Poisson}. The accretion dynamics for Schwarzschild monopole de-Sitter black hole with $Q=0$ has already been discussed in \cite{Mjamil}.

\par The paper is organized as follows: In sec II we derive the metric (line element) for the Reissner-Nordstr\"{o}m anti-de-Sitter (RN-AdS) black hole with global monopole charge. In sec III a general formalism and conservation laws for the accretion process are derived. The speed of sound at the sonic (critical) point is evaluated in sec IV. Next, we have assumed the isothermal equation of state and by choosing different values of state parameter the behaviour of flow for different cases such as ultra-stiff, ultra-relativistic, rotation and sub-relativistic fluids is analysed. In sec VI the general solutions for the polytropic equation of state are obtained as there exists no global (Bondi) solutions for them. In the subsequent section, we have calculated the matter accretion rate of the black hole for the isothermal flow and finally we conclude our discussion. Throughout we use the common relativistic notation while the Greek indices run through $\mu,\nu=0,1,2,3$. The chosen metric signature is $(-,+,+,+)$, and the geometric units as $G=c=1$.

\section{Reissner-Nordstr\"{o}m-(anti)-de Sitter spacetime with global monopole}

We adopted the metric formalism as given in \cite{Naresh}. We assume the general static spherically symmetric metric ansatz of the form
\begin{eqnarray}\label{1}
ds^{2}&=&-f(r)dt^2+\frac{1}{f(r)}dr^{2}+r^2(d\theta^2+\sin^2\theta d\phi^2).
\end{eqnarray}
The Einstein's field equations with the cosmological constant $\Lambda$ are
\begin{eqnarray}\label{2}
R_{\mu \nu}-\frac{1}{2} R g_{\mu \nu}+\Lambda g_{\mu \nu}&=&8\pi \left( T^{(EM)}_{\mu \nu} + T^{(GM)}_{\mu \nu} \right),
\end{eqnarray}
where $T^{(EM)}_{\mu \nu}$ and $T^{(GM)}_{\mu \nu}$ are the energy-momentum tensors for an electromagnetic field and the energy-momentum tensor for the nonminimally coupled global monopole (GM) field, respectively. The field Eqs. (\ref{2}) have non-vacuum solution when the total energy-momentum tensor $T_{\mu \nu} = T^{(EM)}_{\mu \nu} + T^{(GM)}_{\mu \nu}$ does not vanish. The energy momentum tensor for an electromagnetic field is
\begin{eqnarray}\label{3}
T^{(EM)}_{\mu \nu}&=&\frac{1}{4 \pi}(F_{\mu \rho}F_{~\nu}^{\rho}-\frac{1}{4}g_{\mu \nu}F^{\alpha \beta}F_{\alpha \beta}),
\end{eqnarray}
where $F^{\alpha \beta}F_{\alpha \beta} \propto  2 E^2$, $E \propto Q / r^2$, $Q$ is the electric charge and the trace of $T^{(EM)}_{\mu \nu}$ is zero. The Lagrangian which describes a global monopole is given by
\begin{eqnarray}\label{n1}
\mathcal{L}&=&\frac{1}{2}(\partial\psi^{a})^{2}-V(\psi^{a}),
\end{eqnarray}
with $V(\psi^{a})=\frac{\lambda}{4}(\psi^{a}\psi^{a}-\eta^{2})^{2}$ and $\lambda$ is the self-coupling term and $\eta$ is scale of gauge-symmetry breaking $\eta\sim 10^{6}$ GeV \cite{Jusufi}. Here $\psi^{a}$ is a triplet scalar field with $a=1,2,3$ given by
\begin{eqnarray}\label{n2}
\psi^{a}&=&\eta h(r) \frac{x^{a}}{r},
\end{eqnarray}
with $x^{a}x^{a}=r^{2}$ which yields $\psi^2 = \psi^a \psi^a = \eta^2 h(r)^2$. Hence, the energy-momentum tensor for the global monopole field is
\begin{eqnarray}
T^{(GM)}_{\mu \nu}&=&(\partial\psi^{a})^{2} + g_{\mu \nu} \left[\frac{1}{2}(\partial\psi^{a})^{2} -V(\psi^{a}) \right], \label{3-2}
\end{eqnarray}
where $\psi = \pm \sqrt{\psi^a \psi^a}$. Outside the monopole core it must be assumed that $h(r) \rightarrow 1$ as $r \rightarrow \infty$ for $V(\psi^{a})$ to vanish asymptotically. Thus far from the core the stress tensor of the system has the components as
\begin{eqnarray}\label{3-4}
& & T_{~0}^{0\,(GM)} = T_{~1}^{1\,(GM)} = \frac{\eta^2 }{r^2}, \quad  T_{~2}^{2\,(GM)} = T_{~3}^{3\,(GM)} = 0.
\end{eqnarray}
The equation of motion for $\psi^{a}$ is given by
\begin{eqnarray}\label{ra11}
\Box\psi^{a}+\frac{\partial V}{\partial\psi^{a}}=0,
\end{eqnarray}
which upon simplification gives
\begin{eqnarray}\label{ra1}
fh_{,rr}+h_{,r}\Big[f_{,r}+\frac{2}{r}f\Big]-\lambda\eta^{2}h(h^{2}-1)&=&0,
\end{eqnarray}
where $f$ and $h$ are the function of $r$ due to spherical symmetry and is similar to the equation of motion which we obtain through the energy-momentum conservation i.e. $T^{\mu\nu}_{~~;\mu}=0$ \cite{Chen}. A global monopole could readily be added to the scalar field by a general prescription due to Dadhich and Patel \cite{DP} for any spherically symmetric solution.

Using (\ref{3}) and (\ref{3-2}) together with (\ref{3-4}) one can obtain from the Einstein field equations (\ref{2}) that
\begin{eqnarray}\label{4}
& & f(r) =1 + 2 \Phi(r) + \frac{Q^{2}}{r^{2}}-\frac{\Lambda r^{2}}{3},
\end{eqnarray}
where $\Phi(r)$ is the Newtonian gravitational potential and should satisfy the following constraint equations
\begin{eqnarray}
& & \Phi_{,r} + \frac{1}{r} \Phi = \frac{\eta^2}{2 r}, \label{5a} \\
& & \nabla^{2}\Phi = \Phi_{,rr} + \frac{2}{r} \Phi_{,r} = 0. \label{5b}
\end{eqnarray}
Clearly, Eq. (\ref{5b}) is the Cauchy-Euler equation which gives a well known solution
\begin{eqnarray}\label{6}
\Phi(r)&=&\beta-\frac{M}{r},
\end{eqnarray}
where both $\beta = \eta^2 /2$ and $M$ are the constants of integration. These constants are actually the global monopole charge and the mass of the black hole respectively. So using (\ref{6}) the metric function $f(r)$ in (\ref{4}) gets the form
\begin{eqnarray}\label{7}
f(r)&=&1+2\Big(\beta-\frac{M}{r}\Big)+\frac{Q^{2}}{r^{2}}-\frac{\Lambda r^{2}}{3}.
\end{eqnarray}

For $\beta=0$, Eq. (\ref{7}) reduces to the RN-AdS black hole while $\beta=\Lambda=0$ gives Schwarzschild black hole. It is known that a horizon is the null-hypersurface. The normal vector to a constant $r$ hypersurface is $t^{\mu}=(0,1,0,0)$. Clearly $t^{\mu}t_{\mu}=f(r)$ therefore, we can say that $r =const$ is the null-hypersurface at $f(r)=0$ and we can determine all the possible horizons from it. For this, we have a polynomial of degree four by using the Eq. (\ref{7}). When $\Lambda \neq 0$, we get the algebraic equation of the form
\begin{equation} \label{h-eq-2}
r^4 + a_1 r^2 + a_2 r + a_3 = 0,
\end{equation}
where $a_1 = -3 (1+ 2 \beta) / \Lambda, \, a_2 = 6 M / \Lambda$ and $a_3 = -3 Q^2 / \Lambda$. This equation can be solved by making it factorizable such that  $P^2 - R^2= (P + R) (P - R)$ which gives rise to the resolvent cubic equation. Then the quantities $P$ and $R$ in perfect square have the form given by
\begin{equation} \label{h-eq-3}
P = r^2 + x/2, \qquad R = \sqrt{x-a_1} \left( r - \frac{a_2}{2 (x -a_1)} \right),
\end{equation}
if the variable $x$ is chosen such that
\begin{equation} \label{h-eq-4}
x^3 - a_1 x^2 - 4 a_3 x + b = 0,
\end{equation}
i.e. the resolvent cubic with $b = 4 a_1 a_3 - a_2^2 =  36 \left[ (1+ 2\beta) Q^2 - M^2 \right] / \Lambda^2$. Thus, we note that $R$ is linear and $P$ is quadratic in $r$, so each term $P + R$ and $P - R$ is quadratic. Therefore, solving these quadratic formulas one can give all four solutions to the original quartic equation (\ref{h-eq-2}). The cubic equation (\ref{h-eq-4}) can be simplified by making the substitution $x= y + a_1 /3$. In terms of the new variable $y$, Eq.(\ref{h-eq-4}) then becomes $y^3 + 3 L y -2K = 0$, where $K = \left( 2 a_1^3 + 36 a_1 a_3 - 27 b \right) / 54$ and $L = - \left( 12 a_3 + a_1^2 \right) /9$. Now we can solve algebraically the above cubic equation  defining the polynomial discriminant $D=K^2 + L^3$. If $D > 0$, one of the roots is real and the other two roots are complex conjugates. If $D < 0$, all roots are real and unequal. In this case, defining $\theta = \arccos \left( K / \sqrt{-L^3}\right)$, then the real valued solutions of (\ref{h-eq-4}) are of the form
\begin{equation} \label{h-eq-5}
x_{1,2,3} = \frac{a_1}{3} + 2 \sqrt{-L} \cos\left( \frac{2 \pi \ell}{3} + \frac{\theta}{3} \right),
\end{equation}
where $\ell \in \{ 0,1,2 \}$ and $L \leq 0$ which yields $Q^2 \leq (1 + 2 \beta)^2 /(4 \Lambda)$ for $\Lambda > 0$ (de-Sitter spacetime) and $Q^2 \geq (1 + 2 \beta)^2 /(4 \Lambda)$ for $\Lambda < 0$ (anti-de-Sitter spacetime). Thus, the value of $\beta$ for $Q >0$ has to satisfy the relation $ |1 + 2 \beta| > 0$, i.e. $\beta > -1/2$ for both $\Lambda > 0$  and $\Lambda < 0$.  Let $x_1$ be a real root of (\ref{h-eq-4}), the four roots of the original quartic (\ref{h-eq-2}) are given by the roots of the quadratic equations
\begin{equation} \label{h-eq-6}
r^2 \pm \sqrt{x_1 -a_1} \, r + \frac{1}{2} \left( x_1 \mp \frac{a_2}{\sqrt{x_1 -a_1}} \right) = 0,
\end{equation}
which are
\begin{eqnarray} \label{h-eq-7-1}
r_1&=&\frac{1}{2} \left( \sqrt{x_1 -a_1} + \sqrt{\triangle_{-}} \right),\\
r_2&=&\frac{1}{2} \left( \sqrt{x_1 -a_1} - \sqrt{\triangle_{-}} \right),\\
r_3&=&\frac{1}{2} \left( \sqrt{x_1 -a_1} + \sqrt{\triangle_{+}} \right), \\
r_4&=&\frac{1}{2} \left( \sqrt{x_1 -a_1} - \sqrt{\triangle_{+}} \right),  \label{h-eq-7-2}
\end{eqnarray}
where $\triangle_{\pm} = -(x_1 + a_1) \pm 2 a_2 / \sqrt{x_1 - a_1}$ and $x_1 > a_1 = -3 (1+ 2 \beta) / \Lambda$. The polynomial $f=0$ has at most three real and positive roots, which are representing the Cauchy horizon, event horizon and cosmological horizon respectively. In order to find the generic case where these three horizons exist, the values of the parameters $M, Q, \Lambda$ and $\beta$ must be constraint. When all of those parameters are positive and $0 < \Lambda < 0.17$ then three horizons exist, but when $-1 < \Lambda < 0$ and the remaining parameters positive then there exists only two horizons.

Further, from curvature invariants given by Eq. (\ref{e1}), (\ref{e2}) and (\ref{e3}) in Appendix we see that the curvature is finite everywhere outside the horizon and curvature invariants diverge at $r=0$. Therefore, to remove the singularity at horizon we introduce the Eddington-Finkelstein (EF) coordinate \cite{Patryk} which is regular at the event horizon, such that
\begin{eqnarray}\label{10}
dt^{\prime}&=&dt-\frac{2\Big(\frac{M}{r}-\beta\Big)-\frac{Q^{2}}{r^{2}}+\frac{\Lambda r^{2}}{3}}{1+2\big{(\beta-\frac{M}{r})}+\frac{Q^{2}}{r^{2}}-\frac{\Lambda r^{2}}{3}}dr.
\end{eqnarray}
This leads us to the following form of metric:
\begin{eqnarray}\label{11}
& & \fl ds^{2}=-\left[ 1 + 2\left( \beta-\frac{M}{r} \right) + \frac{Q^{2}}{r^{2}}-\frac{\Lambda r^{2}}{3} \right] dt^{\prime2}  -2\left[ 2 \left( \beta-\frac{M}{r}\right) + \frac{Q^{2}}{r^{2}} - \frac{\Lambda r^{2}}{3} \right] dt^{\prime}dr \nonumber \\ && + \left[1-2\left( \beta-\frac{M}{r} \right) - \frac{Q^{2}}{r^{2}}+\frac{\Lambda r^{2}}{3}\right] dr^{2} + r^2(d\theta^2+\sin^2\theta d\phi^2).
\end{eqnarray}
The determinant of the metric defined in (\ref{11}) is $g=-r^{4}\sin^{2}\theta$ and $\sqrt{\mid g\mid}=r^{2}\sin\theta$.

\section{General equations for spherical accretion}

Now we derive the governing equations for spherical accretion. Here, we consider the perfect fluid and analyse the accretion rate and flow of a perfect fluid near RN-AdS black hole with a global monopole charge. For this we define the two basic laws of accretion, i.e. particle conservation and energy conservation. Let $n$ be the number of particles and $u^{\mu}$ be the four velocity of the fluid, then the particle flux will be $J^{\mu}=n u^{\mu}$. From the law of particle number conservation there will be no change in the number of particles, their number remain conserved. In other words, we can say that for this system, the divergence of $4$-vector current density vanishes. Mathematically, it means that
\begin{eqnarray}\label{12}
\nabla_{\mu}J^{\mu}&=&0,
\end{eqnarray}
where $\nabla_{\mu}$ is the covariant derivative. On the other hand, the energy momentum tensor for a perfect fluid is given by $T^{\mu \nu}=(e+p)u^{\mu}u^{\nu}+pg^{\mu \nu}$ with $e$ as the energy density and $p$ as the pressure and its conservation is given by
\begin{eqnarray}\label{13}
\nabla_{\mu}T^{\mu \nu}&=&0.
\end{eqnarray}
The Bondi-type accretion is steady state and spherically symmetric \cite{1t,t1}, so
all the quantities must be function of the radial coordinate only. Furthermore, we are assuming that the fluid is flowing radially in the equatorial plane $(\theta=\frac{\pi}{2})$ therefore, $u^{\theta}=u^{\phi}=0$. By the normalization condition for $4$-velocity $u^{\mu}u_{\mu}=-1$ we obtain,
\begin{eqnarray}\label{14}
u^{t}&=&\frac{\sqrt{ f(r)^2 +(u^{r})^{2}}}{f(r)},
\end{eqnarray}
which yields
\begin{eqnarray}\label{15}
u_{t}&=&-\sqrt{f(r)+(u^{r})^{2}}.
\end{eqnarray}
At the equatorial plane the continuity equation (\ref{12}) becomes
\begin{eqnarray}\label{16}
\nabla_{\mu}(n u^{\mu})&=& \frac{1}{\sqrt{-g}}\partial_{\mu}(\sqrt{-g}n u^{\mu})\nonumber \\&=&
\frac{1}{r^{2}}\partial_{r}(r^{2}n u^{r})=0,
\end{eqnarray}
which upon integration yields
\begin{eqnarray}\label{17}
r^{2}n u^{r}&=&C_{1},
\end{eqnarray}
where $C_{1}$ is a constant of integration. For inward flow $u^{r}<0$ and so $C_{1}<0$. Similarly, we know that enthalpy is the ratio between density and the total internal energy of the system at constant pressure. Therefore, by using the first law of thermodynamics, we can  define enthalpy as $h(e,p,n)=\frac{e+p}{n}$. As the flow is smooth therefore, Eq. (\ref{13}) leads us to
\begin{eqnarray}\label{18}
n u^{\mu}\nabla_{\mu}(hu^{\nu})+g^{\mu\nu}\partial_{\mu}p&=&0.
\end{eqnarray}
Further, we assume that the entropy of a fluid moving along a stream line is constant, so the flow must be isentropic \cite{Pringle}. Hence, Eq. (\ref{18}) gives
\begin{eqnarray}\label{20}
u^{\mu}\nabla_{\mu}(hu_{\nu})+\partial_{\nu}h&=&0.
\end{eqnarray}
Taking the zeroth component of Eq. (\ref{20}) we obtain
\begin{eqnarray}\label{23}
\partial_{r}(hu_{t})=0,
\end{eqnarray}
which upon integration gives
\begin{eqnarray}\label{24}
h\sqrt{f(r)+(u^{r})^{2}}&=&C_{2},
\end{eqnarray}
where $C_2$ is an another constant of integration. Now, these Eqs. (\ref{17}) and (\ref{24}) are the main equations which will be used further to analyse the flow of a perfect fluid in the background of RN-AdS with global monopole.

\section{Sonic points}

Sonic point is the critical point where the four-velocity of the moving fluid becomes equal to the local speed of sound therefore, the flow passing through the sonic point has the maximum accretion rate. If we take constant pressure into account, i.e. $h=h(n)$ then fluid becomes barotropic and the equation of state for barotropic flow can be written as \cite{Ficek}
\begin{eqnarray}\label{33}
\frac{dh}{h}=a^2 \frac{dn}{n},
\end{eqnarray}
where $a$ is the local speed of sound. So Eq. (\ref{33}) yields $\ln h = a^2 \ln n$. From Eqs. (\ref{17}), (\ref{24}) and (\ref{33}), we obtain
\begin{small}
\begin{eqnarray}\label{34}
\left[\left(\frac {u^{r}}{u_{t}}\right)^2-a^2\right] \left( \ln u^{r} \right)_{,r}&=&\frac{1}{r(u_{t})^2}\left[2a^{2}(u_{t})^2 - \frac{1}{2} r f_{,r} \right].~~~~~~
\end{eqnarray}
\end{small}
In this paper, the quantities referring to the critical point will be denoted with the subscripted letter ``$c$". At critical point both sides of Eq. (\ref{34}) must be equal to zero. As $(\ln u^{r})_{,r} \neq 0$ so, the local speed of sound at the sonic point becomes
\begin{eqnarray}\label{35}
a_c^{2} = \left(\frac{u_c^{r}} {u_{t_c}}\right)^{2},
\end{eqnarray}
where $a_c$ is the value of local speed of sound at sonic point, $r_c$ is the distance of fluid from the black hole at sonic point and $u_c^r$ is the velocity of the fluid at the sonic point.
Then, the rhs of Eq. (\ref{34}) at the sonic point given by
\begin{eqnarray}\label{36}
2a_c^{2} (u_{t_c})^{2}-\frac{1}{2} r_c f_{c,r_c} = 0,
\end{eqnarray}
where $f_c = f(r)|_{r=r_c}$, $f_{c,r_c} = f_{,r} |_{r= r_c}$ and $u_{t_c} = u_t (r_{c},u^r_{c})$. Putting Eq. (\ref{35}) into (\ref{36}) we obtain the expression for the radial velocity at sonic point as
\begin{eqnarray}\label{37}
(u_c^{r})^2 &=& \frac{1}{4} r_c  f_{c,r_c} .
\end{eqnarray}
Using Eqs. (\ref{15}), (\ref{36}) and (\ref{37}) we get
\begin{eqnarray}\label{38-1}
r_c f_{c,r_c} = 4 a_c^2 \left[ f_c + (u_c^r)^2 \right], \label{38-2}
\end{eqnarray}
which leads to
\begin{eqnarray}\label{38-4}
a_c^2 &=& \frac{r_c f_{c,r_c}}{r_c f_{c,r_c} + 4 f_c}.
\end{eqnarray}
This equation allows us to determine $r_c$ once the speed of sound $a^2 = dp / de$ is known. Solving Eqs. (\ref{37}) and (\ref{38-4}), we may find the values of $r_c$ and $u_c^{r}$ and so we get the critical point as $(r_c,\pm u_c^{r})$.

\section{Isothermal test fluids}
Isothermal fluids are those fluids which flow at a constant temperature so that the sound speed throughout the accretion process remains constant. As the fluid is flowing at a very fast speed so it does not take the opportunity to exchange the heat with the surroundings, so it is more likely that our dynamical system is adiabatic. The equation of state for such fluids is of the form $p= k e$, where $k$ is the state parameter such that $0 < k \leq 1$ and $e$ is the energy density \cite{t1}. In general, the adiabatic sound speed is defined as $a^{2}= {dp}/{de}$. If we compare it to the equation of state, we find
$a^{2}=k$. Since there is no change in entropy, so $T dS = 0$ where $S$ denotes the entropy. By the first law of thermodynamics
\begin{eqnarray}\label{39}
\frac{de}{dn}&=&\frac{e+p}{n}=h.
\end{eqnarray}
On integrating it from the sonic point to any point inside the fluid,  we obtain
\begin{eqnarray}\label{40}
n&=&n_{c}\exp\left(\int_{e_{c}}^{e}\frac{d e^{\prime}}{e^{\prime} + p \left(e^{\prime}\right)} \right).
\end{eqnarray}
For the isothermal equation of state $p=ke$, Eq. (\ref{40}) becomes
\begin{eqnarray}\label{40a}
n&=&n_{c}\left(\frac{e}{e_{c}}\right)^\frac{1}{k + 1}.
\end{eqnarray}
Comparing this to enthalpy, we get
\begin{eqnarray}\label{41}
h &=& \frac{(k+1)e_{c}}{n_{c}} \left( \frac{n}{n_{c}} \right)^{k},
\end{eqnarray}
By using Eq. (\ref{41}) in Eq. (\ref{24}) we have
\begin{eqnarray}\label{42}
n^{k} \sqrt{ f(r) + (u^r)^2 } &=& C_{3},
\end{eqnarray}
where $C_3 = C_2 n_c^{1-k} / (k+ 1) e_c$. Comparing  Eqs. (\ref{17}) and (\ref{42}), we get
\begin{eqnarray}\label{43}
\sqrt{ f(r) +(u^{r})^2}= C_{3}r^{2k}(u^{r})^{k}.
\end{eqnarray}
On the other hand, we can define the Hamiltonian \cite{A1,A2} as
\begin{eqnarray}\label{41b}
\mathcal{H} &=& \frac{f^{1-k}}{(1-v^{2})^{1-k}v^{2k}r^{4k}},
\end{eqnarray}
where $v$ is the three-dimensional speed for the radial motion in equatorial plane and is defined as $v\equiv \frac{dr}{fdt}$. Consequently, we have
\begin{eqnarray}\label{41bb}
v^{2}=\Big(\frac{u}{fu^{t}}\Big)^{2}=\frac{u^{2}}{u_{t}^{2}}=\frac{u^{2}}{f+u^{2}},
\end{eqnarray}
such that $u^{r}=u=\frac{dr}{d\tau}$, $u^{t}=\frac{dt}{d\tau}$, $u_{t}=-fu^{t}$.
As we are mainly concerned in finding the solutions at the sonic point,
therefore Eqs. (\ref{37}) and (\ref{38-2}) lead to
\begin{eqnarray}\label{44a}
(u_c^{r})^2 &=&  \frac{1}{4}r_c f_{c,r_c} \\ &=&
k \left( \frac{1}{4} r_c f_{c,r_c} + f_c \right). \label{44b}
\end{eqnarray}

It is worthwhile to generalize this discussion from a single point to the continuous flowing fluid. So, we will now analyse the behaviour of fluid by choosing different values of state parameter. For instance, we have $k=1$ (ultra-stiff fluid), $k=1/2$ (ultra-relativistic fluid), $k=1/3$ (radiation fluid) and $k=1/4$ (sub-relativistic fluid).

\subsection{Solution for $k=1$}
\begin{figure}[!ht]
\centering
\includegraphics[width=8cm]{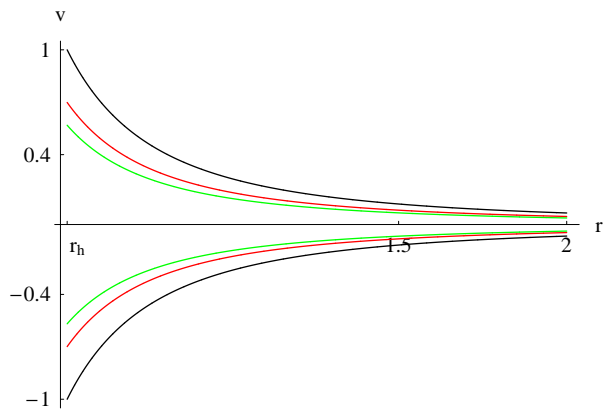}
\caption{Plot showing the trajectories of solutions to Eq. (\ref{41b}) in phase space and the parameters taken as $k=1$, $M=1$, $Q=0.85$, $\beta=0.075$ and $\Lambda=-0.075$. The black curves show the solution at critical point for which $\mathcal{H}=\mathcal{H}_{c}$, the red curves show the solution at $\mathcal{H}=\mathcal{H}_{c}+15.004$ and the green for $\mathcal{H}=\mathcal{H}_{c}+30.007$.}\label{f1}
\end{figure}

For the ultra-stiff fluids, energy density becomes equal to the pressure so that the equation of state is of the form $p = e$, (i.e. $k=1$). From Eqs. (\ref{44a}) and (\ref{44b}) one can find $f_c = 0$, i.e. the expression for $r_c$ is identical to the expression for the locations of event horizon, so the sonic point and the event horizon are located at the same place, i.e. $r_h = r_c$. The Hamiltonian (\ref{41b}) in this case acquires the form:
\begin{eqnarray}\label{46a}
\mathcal{H}&=&\frac{1}{v^{2}r^{4}}.
\end{eqnarray}
The plot in figure \ref{f1} shows the two types of fluid motion. First is the supersonic accretion flow in the upper half, i.e. the region where $v>0$ and the other is the subsonic accretion flow in the lower region where $v<0$.

\subsection{Solution for $k=1/2$}
\begin{figure}[!ht]
\centering
\includegraphics[width=8cm]{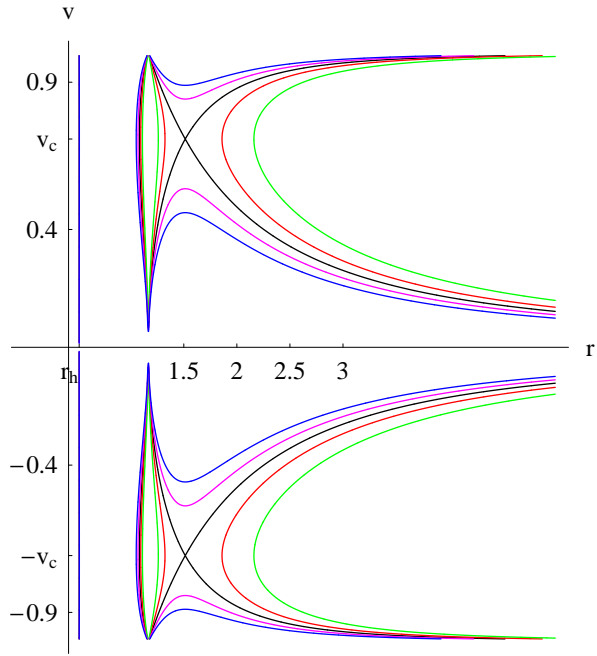}
\caption{The plot shows the trajectories of solutions to Eq. (\ref{41b}) in phase space where the parameters are $k=1/2$, $M=1$, $Q=0.85$, $\beta=0.075$ and $\Lambda=-0.075$. The black curves show the solution for $\mathcal{H}=\mathcal{H}_{c}$, the red curves show the solution for $\mathcal{H}=\mathcal{H}_{c}-0.04$, the green curves show the solution for $\mathcal{H}=\mathcal{H}_{c}-0.09$, the magenta curve is for $\mathcal{H}=\mathcal{H}_{c}+0.04$ and the blue curve is for $\mathcal{H}=\mathcal{H}_{c}+0.09$.}\label{f2}
\end{figure}

For ultra-relativistic fluids the equation of state is of the form $p=e / 2$ (i.e. $k=1/2$). In this case pressure is always less than the energy density. Thus, from Eqs. (\ref{44a}) and (\ref{44b}) we find $r_c f_{c,r_c} - 4 f_c = 0$ which further reduces to the quartic equation
\begin{eqnarray}\label{47}
r_c^{4} + a_1 r_c^{2}+ a_2 r_c + a_3 = 0,
\end{eqnarray}
where $a_1 = - 6 (1+ 2 \beta) / \Lambda, a_2 = 15 M / \Lambda$ and $a_3 = -9 Q^2 / \Lambda$. On further simplification, we may find that the values of $r_c$ are same as given in (\ref{h-eq-7-1}) and (\ref{h-eq-7-2}). Putting this $r_c$ into (\ref{44a}) we get the value $u_c^r$ and then we solve these two equations to get the two critical points as $(r_c,\pm u_c^{r})$.
For instance, the relation between $r$ and $u^{r}$ is obtained from Eq. (\ref{43}) as
\begin{eqnarray}\label{49}
u^{r}=\frac{1}{2}C r^{2}\pm\frac{1}{2} \sqrt{C^{2}r^{4}-4 f(r)},
\end{eqnarray}
where $C= C_3^2$. The Hamiltonian (\ref{41b}) reduces to
\begin{eqnarray}\label{49b}
\mathcal{H}&=&\frac{\sqrt{f}}{r^{2}v\sqrt{1-v^{2}}}.
\end{eqnarray}
The plot in figure \ref{f2} shows the different behaviours for the motion of a fluid. The green and red curves are unphysical since they are double valued. The black curves correspond to the transonic behaviour whereas, the blue and magenta curves show the supersonic behaviour of fluid in the region $v > v_{c}$ and subsonic behaviour for $v < v_{c}$, where $v_c$ is the three-dimensional speed for the radial motion in equatorial plane at the sonic point, which is defined as $v_c = \sqrt{u_c^2 / (f_c + u_c^2)}$ by the Eq. (\ref{41bb}).

\subsection{Solution for Radiation Fluid $(k=1/3)$}
\begin{figure}[!ht]
\centering
\includegraphics[width=8cm]{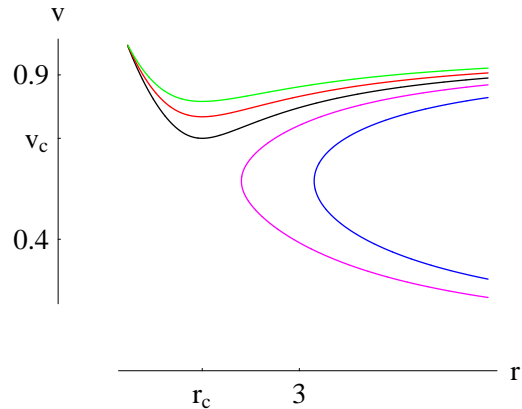}
\caption{The plot showing the trajectories of solutions to Eq. (\ref{41b}) in phase space with the parameters $k=1/3$, $M=1$, $Q=0.85$, $\beta=0.075$ and $\Lambda=-0.075$. The black curve shows the solution for $\mathcal{H}=\mathcal{H}_{c}$,  the red curve shows the solution for $\mathcal{H}=\mathcal{H}_{c}+0.04$, the green curve shows the solution for $\mathcal{H}=\mathcal{H}_{c}+0.09$, the magenta curve is for $\mathcal{H}=\mathcal{H}_{c}-0.04$ and the blue curve is for $\mathcal{H}=\mathcal{H}_{c}-0.09$.}\label{f3}
\end{figure}

The fluid which obeys the equation of state $p=e/3$ (i.e. $k=1/3$) is called radiation fluid. So Eqs. (\ref{44a}) and (\ref{44b}) lead to $r_c f_{c,r_c} - 2 f_c = 0$, i.e.
\begin{eqnarray}\label{50}
(1+2\beta)r_c^{2}- 3 M r_c + 2Q^{2}&=&0,
\end{eqnarray}
which has the solutions
\begin{eqnarray}\label{51}
r_{c\pm} &=& \frac{3M \pm \sqrt{9M^{2}-8Q^{2}(1+2\beta)}}{2},
\end{eqnarray}
where $\beta \leq \left( 3 M / 4 Q \right)^2 -1/2$. Using Eq. (\ref{51}) in (\ref{43}), we obtain
\begin{eqnarray}\label{52}
(u_c^{r})^{2} &=&\frac{[f_c +(u_c^{r})^{2} ]^{3}}{C_3^6 r_c^{4}}.
\end{eqnarray}
From here, we can find the value of $u_c^{r}$, and so we get the critical points $(r_c,\pm u_c^{r})$. Thus, we can express $u^{r}$ in terms of $r$ by Eq. (\ref{44a}).
The Hamiltonian (\ref{41b}) in this case will be
\begin{eqnarray}\label{52a}
\mathcal{H}&=&\frac{f^{2/3}}{r^{4/3}v^{2/3}(1-v^{2})^{2/3}}.
\end{eqnarray}
In figure \ref{f3} we see that fluid's motion is supersonic in the region $v>v_{c}$ where the curves with blue and magenta color shows the unphysical behaviour of the fluid.

\subsection{Solution for $k=1/4$}
When energy density exceeds the isotropic pressure, we get the sub-relativistic fluids and they obey the equation of state $p= e / 4$ (i.e. $k=1/4$). Eqs. (\ref{44a}) and (\ref{44b}) yield $4 f_c-3 r_c f_{c,r_c}=0$ which is a quartic equation and has the following form
\begin{eqnarray}\label{53}
r_c^{4}+ a_1 r_c^2 + a_2 r_c + a_3 =0,
\end{eqnarray}
where $a_1 = 6 (1 + 2 \beta) / \Lambda, \, a_2 = - 21 M / \Lambda$ and $a_3 = 15 Q^2 / \Lambda$. Now Eq. (\ref{53}) has the same form as Eq. (\ref{h-eq-2}) so its roots can be found by following the same procedure as given in Eqs. (\ref{h-eq-7-1}) to (\ref{h-eq-7-2}).
Using the value of $r_c$, we obtain $u_c^{r}$ and similarly $C_3$ from (\ref{43}). Thus, we can express $u_c^{r}$ in terms of $r_c$ by Eq. (\ref{43}) as
\begin{eqnarray}\label{55}
u_c^{r} = \frac{\left[f_c +(u_c^{r})^{2}\right]^{2}}{C_3^{2} r_c^{2}}.
\end{eqnarray}

\begin{figure}[!ht]
\centering
\includegraphics[width=8cm]{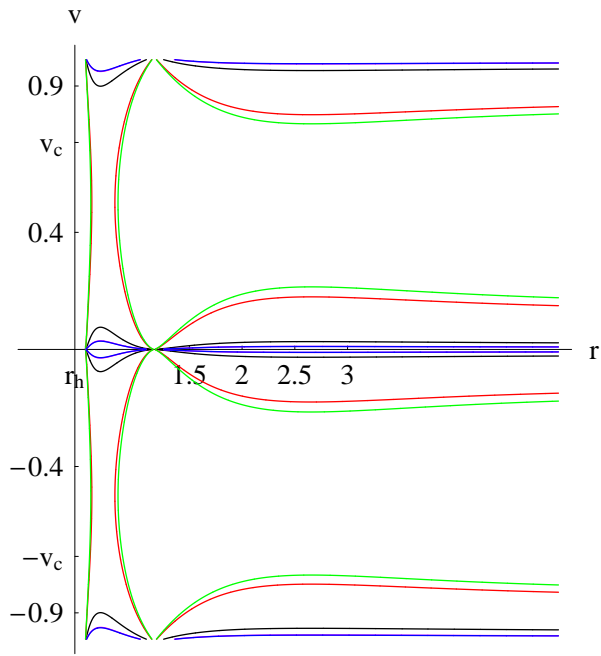}
\caption{The plot shows the trajectories of solutions to Eq. (\ref{41b}) in phase space with parameters as $k=1/4$, $M=1$, $Q=0.85$, $\beta=0.075$ and $\Lambda=-0.075$. The black curve shows the solution for $\mathcal{H}=\mathcal{H}_{c}$, the red curve shows the solution for $\mathcal{H}=\mathcal{H}_{c}-1.04$, the green curve shows the solution for $\mathcal{H}=\mathcal{H}_{c}-1.09$, the magenta curve is for $\mathcal{H}=\mathcal{H}_{c}+1.04$ and the blue curve is for $\mathcal{H}=\mathcal{H}_{c}+1.09$.}\label{f4}
\end{figure}

By repeating the same procedure as in previous cases again we can find an explicit form of solution and critical points $(r_c,\pm u_c^{r})$. The Hamiltonian (\ref{41b}) takes the form
\begin{eqnarray}\label{55a}
\mathcal{H}&=&\frac{f^{3/4}}{rv^{1/2}(1-v^{2})^{3/4}}.
\end{eqnarray}
Figure \ref{f4} shows the motion of the fluid in different regions. The motion of the fluid is supersonic in the region where $v>v_{c}$ and subsonic where $v<v_{c}$ whereas, the region of vertical curves show unphysical behaviour of the flow.

\par
\begin{figure}[!ht]
\centering
\includegraphics[width=10cm]{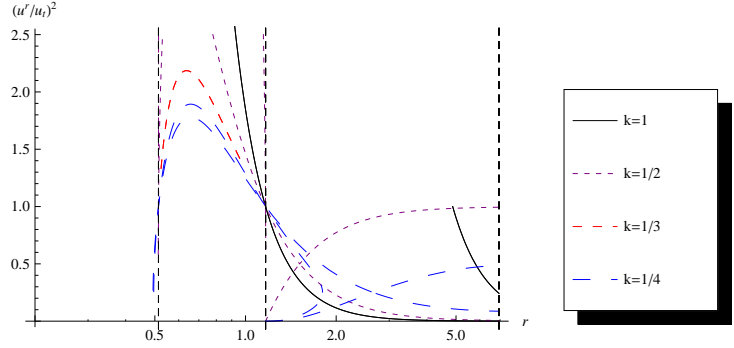}
\caption{Transonic solutions obtained for the isothermal equation of state $p=ke$. The other parameters are taken as $M=1$, $\beta=0.075$, $Q=0.85$ and $\Lambda=-0.075$.}\label{f6}
\end{figure}

 In the above cases we have discussed the non-transonic solutions but we are also concerned with the flows which pass through the transonic point. Transonic flows are often considered in a context of spherical accretion which yields to the maximum accretion rate. So, in figure \ref{f6} we have shown the transonic behaviour of the flow for the above mentioned fluids. Since transonic solutions form a closed orbit therefore, the figure demonstrates that it may possible to get homoclinic orbits around a black hole in case of subsonic solutions.

\section{ Polytropic test fluids}
Polytropes are self-gravitating gaseous spheres that are, very useful as crude approximation to relativistic fluid models. The term ``polytropic" was originally used to describe a reversible process on any open or closed system of gas or any fluid which involves both heat and work transfer, such that a specified combination of properties were maintained constant throughout the process. In such a process, the expression relating the properties of the system throughout the process is called the polytropic path. There are an infinite number of reversible polytropic paths between two given states; the most commonly used polytropic path is $TdS=C$, where $T$ is temperature, $S$ is entropy, and $C$ is an arbitrary constant which is equal to zero for an adiabatic process.
This path is equivalent to the assumption that the same amount of heat is transferred to the system in each equal temperature increment. The equation of state, which the fluid has when it follows this path is called the \emph{Polytropic Equation of State}. Mathematically, it is given as $p=kn^{\Gamma}$, where $k$ and $\Gamma$ are the constants. By keeping this equation of state in mind, we get the following expression for enthalpy \cite{ggm}
\begin{eqnarray}\label{66}
h=\frac{\Gamma-1}{\Gamma-1-a^{2}}.
\end{eqnarray}
Using Eq. (\ref{66}),  Eq. (\ref{24}) has taken the form
\begin{eqnarray}\label{67}
\frac{\sqrt{ f(r)+(u^{r})^{2}}}{\Gamma-1-a^{2}}=C_6,
\end{eqnarray}
where $C_{6}$ is the arbitrary constant. If we take our boundary at infinity (i.e $r = r_{\infty}$) then Eq. (\ref{67}) will be
\begin{eqnarray}\label{68}
\frac{\sqrt{ f_{\infty} +(u^{r}_{\infty})^{2}}}{\Gamma-1-a^{2}_{\infty}}=C_6,
\end{eqnarray}
where $f_{\infty} = f(r_{\infty})$. On comparing Eq. (\ref{67}) and Eq. (\ref{68}), we obtain
\begin{eqnarray}\label{69}
(\Gamma-1-a^{2}_{\infty})\sqrt{ f(r)+(u^{r})^{2}}=&&(\Gamma-1-a^{2}) \sqrt{ f_{\infty} +(u^{r}_{\infty})^{2}}.
\end{eqnarray}
Now at the sonic point, the enthalpy can be given as
\begin{eqnarray}\label{70}
\frac{h}{h_{\infty}}=\left(\frac{n}{n_{\infty}}\frac{a^{2}_{\infty}}{a^{2}}\right)^{\Gamma-1},
\end{eqnarray}
which further on using Eq. (\ref{66}) gives rise to
\begin{eqnarray}\label{71}
n=n_{\infty}\left(\frac {a^{2}}{a^{2}_{\infty}}\frac{\Gamma-1-a^{2}}{\Gamma-1-a^{2}_{\infty}}\right)^{\frac{1}{\Gamma-1}}.
\end{eqnarray}

Since $r^{2}n u^{r}= C_1$, at the spacial infinity it will be $r^{2}_{\infty}n_{\infty}u^{r}_{\infty}= C_1$.
On combining Eqs. (\ref{70}) and (\ref{71}) we obtain
\begin{eqnarray}\label{72}
u^{r}=u^{r}_{\infty} \left(\frac{r_{\infty}}{r} \right)^2 \left[
\left( \frac{a_{\infty}}{a} \right)^2 \left( \frac{\Gamma-1-a^{2}}{\Gamma-1-a^{2}_{\infty}} \right) \right]^{\frac{1}{\Gamma-1}}.
\end{eqnarray}
At the critical point $r_c$, together with (\ref{37}) and (\ref{72}), Eq. (\ref{69}) becomes
\begin{eqnarray}\label{73}
(\Gamma-1-a_c^{2})^{2}\left( f_{\infty} + B_3 \right)=&&(\Gamma-1-a^{2}_{\infty})^{2} \left( f_c + \frac{1}{4} r_c f_{c,r_c} \right),
\end{eqnarray}
where
\begin{eqnarray}\label{74}
B_3 &\equiv& \left(u_c^{r} \right)^{2}\frac{r_c^{4}}{r_{\infty}^{4}}\left(\frac{a_c^{2}}{a^{2}_{\infty}}
\frac{\Gamma-1-a^{2}_{\infty}}{\Gamma-1-a_c^{2}}\right)^{\frac{2}{\Gamma-1}}.
\end{eqnarray}
Note that Eq. (\ref{73}) has only one unknown $r_c$. So, if we solve this equation with boundary values of $r_{\infty}$ and $a_{\infty}$, we obtain the position of critical values of $r_c$, $a_c^{2}$ and $u_c^{r}$. Similarly, as in isothermal case we can compute two critical points $(r_c,\pm u_c^{r})$ in the polytropic case also. Furthermore, Eq. (\ref{72}) can also be solved numerically to obtain the function  $u^{r}$.

\section{Black hole's accretion rate}
The rate of change in the mass of a black hole is called mass accretion rate and is generally represented by $\dot M$. Basically, it measures the mass of a black hole per unit time. It is defined as the area times flux of a black hole at the event horizon. In this section, we will discuss the effect of radius on the accretion rate. The general expression to calculate it is $\dot M\left|_{r_{h}}=4\pi r^{2} T^{r}_{t}\right|_{r_{h}}$ \cite{Mjamil}, which refers to relativistic statement of the flux of mass-energy density. In non-relativistic treatment, $\dot{M}$ is defined as the flux of rest-mass density \emph{or} $\dot{M} = 4 \pi r^2 \rho u^r$, where $\rho$ is the rest-mass density, $\rho = m_b\, n_b$, $m_b$ is the average mass per baryon and $n_b$ is the baryon density. Then it follows form the perfect fluid energy-momentum tensor (\ref{3}) that $T_t^r = (e + p) u_t u^r$ \cite{M1}. As our dynamical system is conserved so we have $\nabla_{\mu}J^{\mu}=0$ and $\nabla_{\nu} T^{\mu \nu}=0$. From these conservation equations defined in Eqs. (\ref{17}) and (\ref{24}) we obtain
\begin{eqnarray}\label{56}
& & r^2 u^{r} \, (e + p)  \sqrt{f(r) + (u^{r})^{2}} = A_0,
\end{eqnarray}
where $A_0$ is an arbitrary constant. Now assuming the equation of state $p=p(e)$, the relativistic energy flux (or continuity) equation gives
\begin{eqnarray}\label{57}
& &  \frac{de}{e + p} + \frac{du^{r}}{u^{r}} + \frac{2}{r} dr = 0.
\end{eqnarray}
Integration of this equation yields
\begin{eqnarray}\label{58}
& & r^2 u^{r} \, \exp \left[ \int_{e_{\infty}}^{e} \frac{d e'}{e' + p(e')}  \right] = - A_1,
\end{eqnarray}
where $A_1$ is an integration constant, $e_{\infty}$ is the matter density at infinity and the minus sign is taken because $u^{r}<0$. If we combine the Eq. (\ref{58}) with (\ref{56}) we obtain
\begin{eqnarray}\label{59}
A_3 = - \frac{A_0}{A_1} =&&(e + p)\sqrt{f(r) + (u^{r})^2} \exp \left[ -\int_{e_{\infty}}^{e} \frac{d e'}{e' + p(e')} \right],
\end{eqnarray}
where $A_3$ is an arbitrary constant. If we take the boundary condition at infinity, then the constant $A_3$ becomes $A_3=e_{\infty}+p(e_{\infty})=- A_0/A_1$, where $A_0=(e + p)u_t u^{r} r^2 =- A_1(e_{\infty}+p(e_{\infty}))$. Furthermore, on equatorial plane due to the spherical symmetry, the equation of mass flux  $\nabla_{\mu} J^{\mu} = 0$ can also be written as
\begin{eqnarray}\label{60}
& & r^2 u^{r}\, n = A_2,
\end{eqnarray}
where $A_2$ is an integration constant. If we divide the Eq. (\ref{56}) with (\ref{60}), we get an another useful relation
\begin{eqnarray}\label{60-2}
& & \frac{e + p}{n} \sqrt{f(r) + (u^{r})^2} = \frac{A_0}{A_2} \equiv  A_4,
\end{eqnarray}
where $A_4$ is a constant such that $A_4 = \left( e_{\infty}+p_{\infty}\right) / n_{\infty}$ \cite{Babichev}. Using (\ref{56}), the black hole's rate of change of mass takes the following form
\begin{eqnarray}\label{61}
\dot{M} &=& - 4 \pi r^2 u^{r} (e + p) \sqrt{f(r) + (u^{r})^2 } =   - 4 \pi A_0.
\end{eqnarray}
Then, it becomes
\begin{eqnarray}\label{61-1}
\dot{M} &=&  4\pi A_1 (e_{\infty} + p(e_{\infty})),
\end{eqnarray}
due to the boundary condition at infinity. This result is valid for any fluid which obeys the equation of state of the form $p = p(e)$. So, the accretion rate for the black hole will be
\begin{eqnarray}\label{61-2}
\dot{M} &=& 4\pi A_1 (e + p) |_{r=r_{h}} ,
\end{eqnarray}
at black hole horizon $r_h$. Thus, evaluating Eq.(\ref{61-2}) at $r_h$, one can obtain the black hole mass rate for observers at the black hole horizon.

Let us take an isothermal equation of state, i.e. $p=k e$, which implies that $(e+p)=e(1+k)$. Then Eq.(\ref{58}) reduces to $r^2 u^{r} e^{\frac{1}{1+k}} = -A_1$, that is
\begin{eqnarray}\label{61-3}
& & e=\left[- \frac{A_1}{ r^2 u^{r} } \right]^{1+k}.
\end{eqnarray}
With this expression of $e$, Eq. (\ref{56}) yields
\begin{eqnarray}\label{61-4}
& & (u^{r})^2 - \frac{A_0^2 A_1^{-2(1+k)} }{(1+k)^2} r^{4 k} \left( -u^{r} \right)^{2k} + f(r) = 0,
\end{eqnarray}
in which the $u^{r}$ can be obtained for the given values of $k$ if the above algebraic equation is solved. Using the obtained $u^{r}$, we find the energy density $e(r)$ from  (\ref{61-3}). For instance, when $k=1$ (ultra-stiff fluids) it follows from Eq.(\ref{61-4}) that
\begin{eqnarray}\label{61-5}
& & u^{r}= \pm 2 A_1^2 \sqrt{ \frac{f(r)}{{A_0^2 r^4 - 4 A_1^4 }} },
\end{eqnarray}
which gives
\begin{eqnarray}\label{61-6}
& & e= \frac{\left( A_0^2 r^4 - 4 A_1^4 \right)}{4 A_1^2 r^4 f(r) }.
\end{eqnarray}
Putting (\ref{61-6}) into (\ref{61-2}), we get
\begin{eqnarray}\label{61-7}
& & \dot{M} = \frac{ 6 \pi  \left( A_0^2 r^4 - 4 A_1^4  \right)}{ A_1 r^2 \left[ 3 Q^2 - 6 M r + 3(1+ 2 \beta) r^2 - \Lambda r^4  \right]}.~~~~~
\end{eqnarray}

\begin{figure}[!ht]
  \centering
\minipage{0.45\textwidth}
 \includegraphics[width=5.9cm,height=5.4cm]{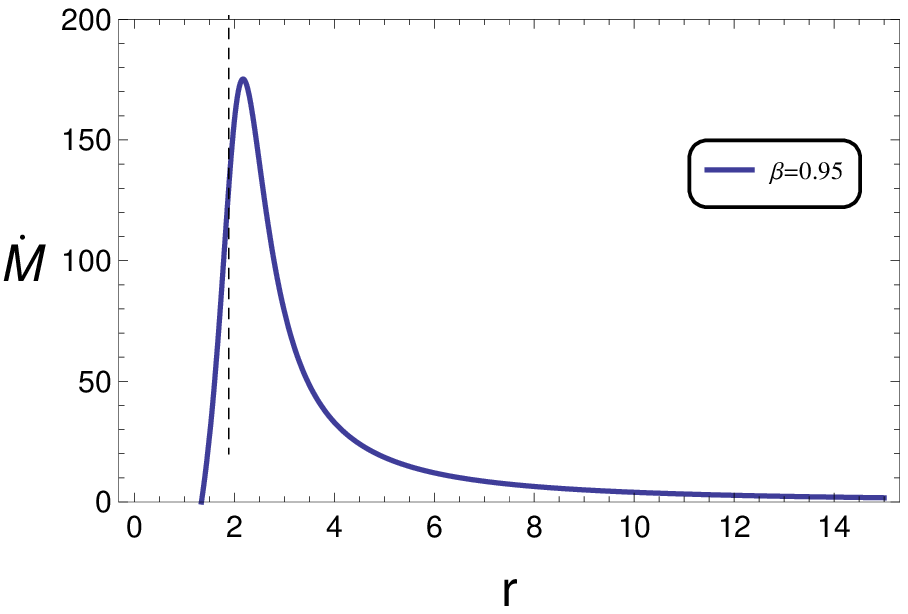}
 \label{1}
\endminipage\hfill
\minipage{0.45\textwidth}
\includegraphics[width=5.9cm,height=5.4cm]{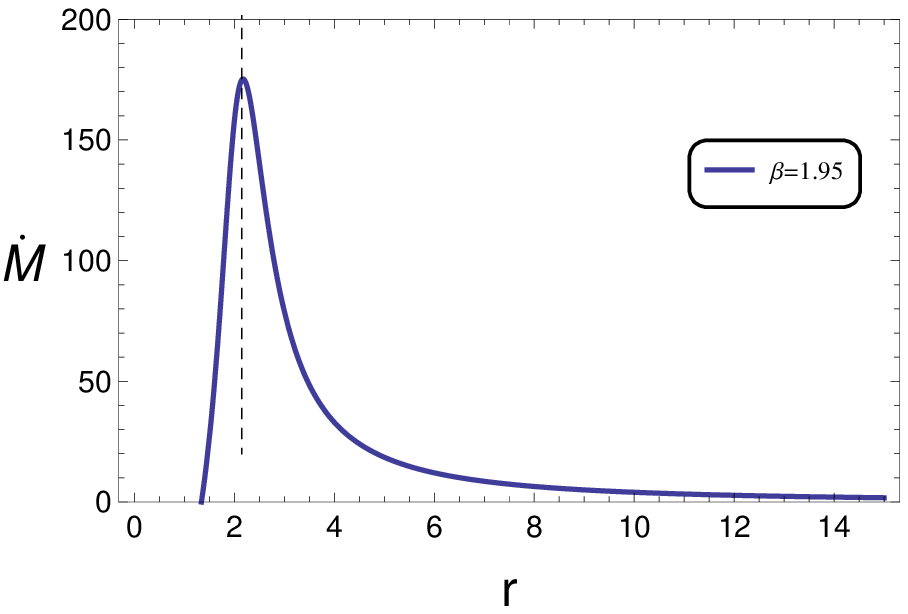}
\label{2}
\endminipage\hfill\\
\minipage{0.45\textwidth}
 \includegraphics[width=5.9cm,height=5.4cm]{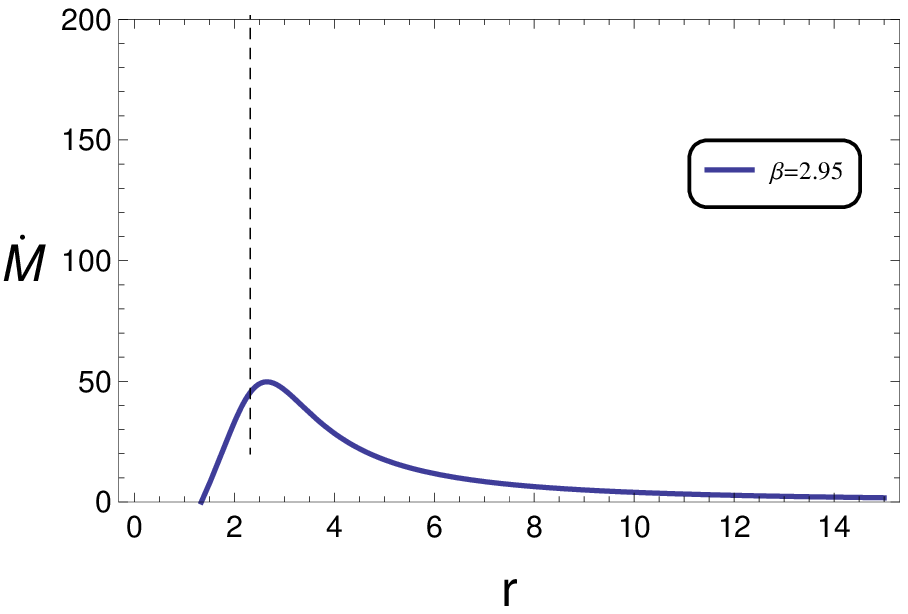}
 \label{1}
\endminipage\hfill
\minipage{0.45\textwidth}
 \includegraphics[width=5.9cm,height=5.4cm]{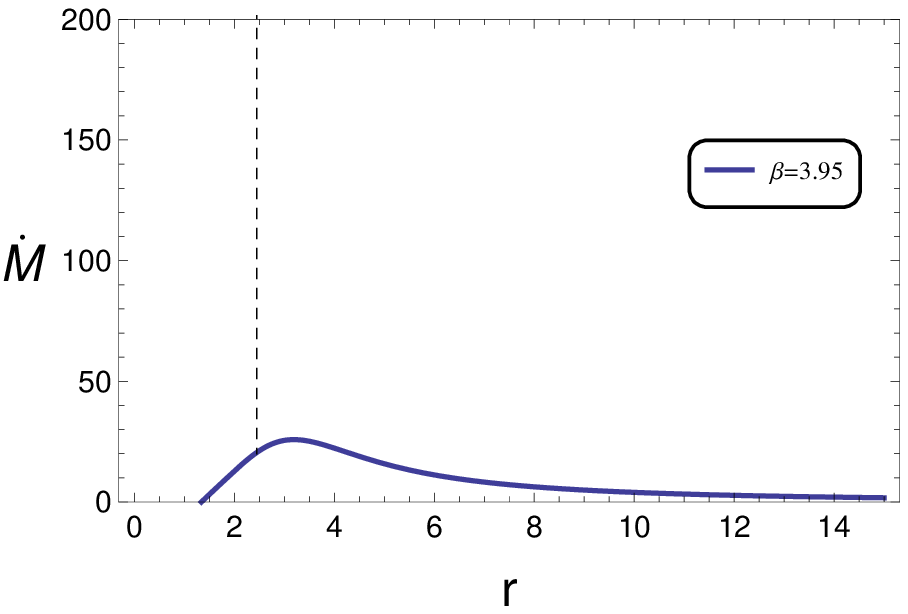}
 \label{1}
\endminipage\hfill
\caption{Plot showing the mass accretion rate for different values of monopole parameter for the ultra-stiff fluid $(k=1)$. The other parameters are fixed and are taken as $M=1$, $\Lambda=-0.2$, $Q=0.75$, $A_{0}=2.5$ and $A_{1}=1$ whereas, the vertical line shows the location of horizon corresponding to monopole parameter in each case.}\label{f5}
\end{figure}
In figure \ref{f5} we plot the mass accretion rate for different values of $\beta$. It can be seen that for all values of $\beta$ the accretion rate increases but when it reaches to the maximum limit, it will decrease. Also, we see the large monopole parameter $\beta$ greatly suppresses the maximum accretion rate.

\section{Discussion}
In $1989$, the first solution of the Einstein field equations with a global monopole was derived \cite{Bar}. Here, we have derived the line element (metric) for the Reissner-Nordstr\"om-(anti)-de-Sitter spacetime with global monopole. By assuming certain conservation laws for the adiabatic system, we explored the behaviour of fluid. Generally, the equation of state helps us to identify about what kind of fluid is accreting onto black hole. Here, we have focussed mainly on the ultra-stiff, ultra-relativistic, radiation and sub-relativistic fluids. Different kind of fluids with the unique value of state parameter have different kind of evolution onto the black hole. We have not considered the cases for the vacuum energy or cosmological constant since their accretion does not change that much evolution of the black hole. Further, we have assumed that only a single test fluid accretes onto black hole at a certain time and discussed the isothermal as well as the polytropic flows. Though we have derived the general expressions for all type of fluids satisfying the isothermal and polytropic equation of state but for isothermal equation of state we have also find the solutions of the mentioned fluids at the sonic point. However, in case of polytropic fluids there exists no global solutions \cite{PatrykMach}. Furthermore, we have determined the general analytical expression for the mass accretion rate $\dot{M}$ and found that $\dot{M}$ depends on the mass and charge of the black hole and large monopole parameter $\beta$ suppresses the maximum accretion rate frequently.

\par A number of extensions to our study is possible. For instance, one may attempt to consider a non-adiabatic system, so that the fluids may have the effects due to viscosity and energy transfer. It is also interesting to extend this result to the region between the sonic point and boundary. A similar analysis can be done for the spherically symmetric but non-static geometry of the fluid where the velocity and energy density of the fluid varies with the time and position.

\section*{Acknowledgements}

UC was supported by The Scientific Research Projects Coordination Unit of Akdeniz University (BAP). We would like to thank the referees for giving insightful comments to improve this work.

\section*{Appendix: Curvature Invariants for the RN-AdS Monopole Black Hole}
\renewcommand{\theequation}{A.\arabic{equation}}
\setcounter{equation}{0}

The curvature invariants are given by
\begin{eqnarray}
& & \fl I_{1}= g^{\mu\nu}R_{\mu\nu}=4\Big(\frac{\beta}{r^2}-\Lambda\Big),\label{e1} \\& & \fl I_{2}= R^{\mu\nu}R_{\mu\nu}=\frac{4(Q^4-2Q^{2}r^{2}\beta+2r^{4}\beta^{2}-2r^{6}\beta\Lambda+r^{8}\Lambda^{2})}{r^8},\label{e2} \\& & \fl I_{3}= R^{\mu\nu\rho\sigma}R_{\mu\nu\rho\sigma} \nonumber \\& & \fl \,\,\,\, =\frac{8\Big(21Q^{4}+6Q^{2}r(-6M+r\beta)+r^{2}(18M^{2}-12Mr\beta+6r^{2}\beta^{2}-2r^{4}\beta\Lambda+r^{6}\Lambda^{2})\Big)}{3r^{8}}.\nonumber\\
~~~\label{e3}
\end{eqnarray}
The above curvature invariants are well defined everywhere except at $r=0$.

\end{document}